\documentclass[aps,prd,reprint,groupedaddress]{revtex4-2}

\usepackage{amsmath,amssymb,amsfonts,euscript}
\usepackage{graphicx}
\usepackage{dcolumn}
\usepackage{bm}

\DeclareMathOperator{\sgn}{sgn}
\DeclareMathOperator{\res}{res}

\newcommand{\der}[2]{\frac{\mathrm{d}#1}{\mathrm{d}#2}}

\newcommand{\pd}[2]{\frac{\partial #1}{\partial #2}}

\begin{document}

\title{Electromagnetic field of a charge asymptotically approaching spherically symmetric black hole}


\author{S. O. Komarov$^{1,2}$}
\email{StasKomarov@tut.by}

\author{A. K. Gorbatsievich$^{1}$}
\email{Gorbatsievich@bsu.by}

\author{G. V. Vereshchagin$^{2,3,4,5}$}
\email{Veresh@icra.it}
\affiliation{$^{1}$Department of Theoretical Physics and Astrophysics, Belarusian State University, 4 Nezavisimosti Ave., 220030 Minsk, Belarus}
\affiliation{$^{2}$ICRANet-Minsk, National Academy of Sciences of Belarus, 68-2 Nezavisimosti Ave., 220072 Minsk, Belarus}
\affiliation{$^{3}$ICRANet, Piazza della Repubblica, 10, 65122 Pescara, Italy}
\affiliation{$^{4}$ICRA, Dipartimento di Fisica, Sapienza Universit\`a di Roma, Piazzale Aldo Moro 5, I-00185 Rome, Italy}
\affiliation{$^{5}$INAF — IAPS, Via del Fosso del Cavaliere, 100, 00133 Rome, Italy}

\date{\today}

\begin{abstract}
	We consider a test charged particle falling onto a Schwarzschild black hole and evaluate its electromagnetic field. The Regge-Wheeler equation is solved analytically by approximating the potential barrier with Dirac delta function and rectangular barrier. We show that for asymptotically large time measured by a distant observer the electromagnetic field approaches the spherically symmetric electrostatic field exponentially fast. This implies that in the region accessible to a distant observer the initial state of separated charge and Schwarzschild black hole becomes asymptotically indistinguishable from the Reisnner-Nordstr\"om solution. Implications of this result for models with plasma accretion on black holes are discussed.
\end{abstract}


\maketitle

\section{Introduction}

Electromagnetic radiation from a charge falling onto a black hole has been subject of extensive research in the 70s \cite{Ross1971,Tiomno1972,Ruffini1972}. The invent of the Regge-Wheeler formalism describing small perturbations of a black hole \cite{1957PhRv..108.1063R} and its extension by Zerilli \cite{1970PhRvD...2.2141Z} allowed the formulation of the basic equations for the problem of electromagnetic radiation of test particle in the field of a spherically symmetric black hole \cite{1972NCimL...3..211R}. These equations were solved numerically and the spectrum of first few multipoles together with total radiated energy were found \cite{Tiomno1972,Ruffini1972}. Later this problem has been addressed in \cite{2003PhRvD..68h4011C} for a particle projected to a black hole with initial velocity, as well as in 
\cite{Shatski-NovikovEN}, 
where also dipole infall was considered. The complex angular momentum technique was used recently \cite{2020PhRvD.102b4026F} to compute radiation spectra for more multipoles than in original papers.

While basic results for the total radiated energy as well as the spectrum of radiation of charged particle in the vicinity of black hole have been established, the question remains about the fate of the electromagnetic field of the charged particle approaching the horizon of the black hole. As observed from infinity, electromagnetic radiation of a charged particle must vanish for asymptotically large times due to the gravitational redshift. Therefore in this limit only static electromagnetic field should remain. Analysis of electromagnetic field of a charge kept at rest near the black hole (see, e.\,g. \cite{Hanni1973,1976JPhA....9.1081L,ThorneMacdonald,1973blho.conf..451R}) shows that the field approaches spherical symmetry when the charge approaches black hole event horizon. In particular, the results from \cite{Hanni1973} imply that higher harmonics become subdominant with respect to the monopole for a charge approaching the black hole. In addition the analysis of the electromagnetic field of such a static charge given analytically in \cite{1976JPhA....9.1081L} shows that the field is spherically symmetric when the charge is located at the event horizon of the black hole. These considerations lead \cite{ThorneMacdonald}, p.50 to conclude that "The electric field of a dynamically infalling particle behaves similarly to this sequence of static fields. Although the particle never, in any finite universal time $t$, goes through the true horizon, soon after it passes the stretched horizon its field behaves as though its electric charge had been deposited on and smeared uniformly over the stretched horizon". Similar conclusion was reached previously in \cite{1973blho.conf..451R}.

It is obvious that the particle cannot be at rest in the vicinity of black hole without any external force, that should also influence the gravitational field of black hole. The goal of the present paper is to consider more physical situation when the charge is not fixed. Here we report the results obtained for the \emph{dynamical} problem, namely radial infall of a particle in Schwarzschild geometry. We find the approximate analytical expressions for electromagnetic field and demonstrate that all multipoles except for the monopole vanish asymptotically for large time measured by a distant observer.

The paper is organized as follows. In Section \ref{sec2} multipole expansion of electromagnetic field is recalled. In Section \ref{sec3} the monopole component of electromagnetic field is considered. In Section \ref{sec4} equations for electromagnetic field are solved analytically by approximating the potential barrier of the Regge-Wheeler equation with Dirac delta function. In Section \ref{sec5} the approximation with a rectangular potential barrier is considered and electromagnetic field is determined in the asymptotic limit of the charge approaching the black hole event horizon. We discuss obtained results in the Discussion Section. 

\section{Multipole expansion of electromagnetic field}
\label{sec2}
Consider a point-like, test particle with electric charge $q$ and mass $m$ moving in the neighbourhood of Schwarzschild black hole. Equations of motion for the particle have the following form (see, e. g. \cite{Mi})
\begin{equation}\label{eqmotion}
m\frac{Du^i}{D\tau}=q F^{ij}u_j\,,
\end{equation} 
where $u^i$ --- 4-velocity vector of the particle, $\tau$ its proper time, $F^{ij}$ --- tensor of electromagnetic field.  The notation $D w^i/D\tau=w^i{}_{;j}u^j$ is used, where $w^{i}{}_{;j}$ denotes covariant derivative of an arbitrary vector field $w^i$ with respect to coordinate $x^j$. We use the system of units in which $c=1$ \footnote{We use the system of units where the speed of light in vacuum is $c=1$. Signature of the metric is +2, $x^0=t$. Greek indices run from 1 to 3, Latin indices run from 0 to 4.}. In the general case $F^{ij}$ represents superposition of external electromagnetic field and electromagnetic field, created by the particle. Hence it accounts also for the radiation reaction.


Electromagnetic field tensor $F^{ij}$ satisfies Maxwell equations in curved space-time (see, e. g. \cite{Mi})
\begin{eqnarray}\label{Maxwell}
&&\left\{
\begin{split}
&F^{ls}{}_{;s}=4\pi j^l\,,\\
&F_{[ij,k]}=0\,,
\end{split}
\right.
\end{eqnarray}
where $j^l$ is the vector of 4-current density.

It follows from (\ref{Maxwell}), that $F_{ij}$ can be represented in the form (see, e. g. \cite{Mi})
\begin{eqnarray}\label{potentials}
&F_{ij}=A_{j;i}-A_{i;j}\,;
\end{eqnarray}
where $A_i$ is the 4-potential of electromagnetic field. In the case of a point like charged particle $j^l$ can be represented as
\begin{eqnarray}\label{current}
&j^l(x^k)\nonumber\\
&=q\dfrac{\delta(x^1-\tilde{x}^1(x^0))\delta(x^2-\tilde{x}^2(x^0))\delta(x^3-\tilde{x}^3(x^0))}{\sqrt{-g}u^4(x^0,\tilde{x}^{\alpha}(x^0))}\nonumber\\
&\times u^l(x^0,\tilde{x}^{\alpha}(x^0))\,.
\end{eqnarray}  
Here $g=\det(g_{ij})$. Coordinates with tilde  $\tilde{x}^i$ are functions describing the world line of the source and without tilde $x^i$ are arbitrary coordinates of space-time describing the point of observation of the vector field $j^l(x^k)$.

Consider the particle moving radially  in the vicinity of Schwarzschild black hole. In Schwarzschild coordinates $\{t,\,r,\,\theta,\,\phi\}$ the metric of space-time has the following form
\begin{equation}
\mathrm{d}s^2=-\left(1-\frac{2M}{r}\right)\mathrm{d}t^2+\frac{\mathrm{d}r^2}{\left(1-\frac{2M}{r}\right)}+r^2\mathrm{d}\theta^2+r^2\sin^2\theta\mathrm{d}\phi^2\,.  
\end{equation}

We choose orientation of the coordinate system in such a way that trajectory of the particle satisfies the condition $\theta=0$. Due to axial symmetry of the describing system, electromagnetic field does not depend on azimuthal angle $\phi$. As it has been shown in \cite{1972NCimL...3..211R}, in this case electromagnetic field of the particle can be represented as
\begin{eqnarray}\label{anzats}
\left\{
\begin{split}
&A_0=\sum\limits_{l=0}^{\infty}f_{l}(r,t)P_l(\cos\theta)\,,\\
&A_1=\sum\limits_{l=0}^{\infty}h_{l}(r,t)P_l(\cos\theta)\,,\\
&A_2=\sum\limits_{l=0}^{\infty}k_{l}(r,t)\der{P_l(\cos\theta)}{\theta}\,,\,A_3=0\,.
\end{split}
\right.
\end{eqnarray}
Here $P_l(x)$ are Legendre polynomials \footnote{For arbitrary motion the electromagnetic field decomposition contains spherical polynomials, which are reduced to Legendre polynomials for radial motion.} with multipole index $l$.

Introduce multipole coefficients $b_{l}(r,t)$ by the following equations
\begin{eqnarray}
&&b_{l}(r,t)=\frac{r^2}{l(l+1)}\left(\pd{h_l}{t}-\pd{f_l}{r}\right)\,,\text{ for }l>0\,;\label{bl}\\
&&b_{0}(r,t)=r^2\left(\pd{h_0}{t}-\pd{f_0}{r}\right)\,.\label{bl0}
\end{eqnarray}
From (\ref{bl}) and (\ref{bl0}) it follows that
\begin{eqnarray}\label{Fl}
&&F_{10}=\frac{1}{r^2}b_0+\sum\limits_{l=1}^{\infty}\frac{l(l+1)}{r^2}b_l(r,t) P_l(\cos\theta)\,.
\end{eqnarray}
It is convenient to introduce the tortoise coordinate $r^*$, which can be defined in the region outside the black hole (see, e. g. \cite{Mi}) as
\begin{equation*}
r^*(r)=r+2M\ln{(r/2M-1)}\,;\quad 
\frac{\mathrm{d}r}{\mathrm{d}r^*}=1-2M/r\,.
\end{equation*}
Let $\tilde b_l(r^*,\omega)$ be the Fourier transform of $b_l(r(r^*),t)$. Substituting (\ref{current}), (\ref{anzats}), (\ref{bl}), (\ref{bl0}) into (\ref{Maxwell}), the following equation is obtained for Fourier components \cite{Tiomno1972,Ruffini1972,1972NCimL...3..211R}
\begin{equation}\label{SchAmp}
\tilde{b}_l''+\left[\omega^2-U(r^*)\right]\tilde{b}_l=e^{i\omega T(r)}a_l(r^*)\,,
\end{equation}
where
\begin{eqnarray}\label{a}
&U(r^*)=\left(1-\frac{2M}{r(r^*)}\right)\frac{l(l+1)}{r(r^*)^2}\,, \\
&a_l(r^*)=\frac{q}{2\pi}\frac{2l+1}{l(l+1)}\left[T''(r^*)+i\omega(T'(r^*))^2-i\omega\right]\,, \\
&a_0(r^*)=\frac{q}{2\pi}\left[T''(r^*)+i\omega(T'(r^*))^2-i\omega\right]\,.
\end{eqnarray}
Here the prime denotes derivative with respect to tortoise coordinate $r^*$; $T(r^*)=t$ represents the equation of the worldline of the particle in arbitrary electromagnetic field, allowing radial motion. For instance, when particle starts at the spatial infinity in absence of electromagnetic fields and without initial velocity, it has the form
\begin{eqnarray}
&T(r^*)=-\frac{2}{3\sqrt{2M}}r(r^*)^{3/2}-2\sqrt{2Mr(r^*)}\nonumber\\
&-2M\ln{\left(\frac{\sqrt{r(r^*)}-\sqrt{2M}}{\sqrt{r(r^*)}+\sqrt{2M}}\right)}\,. \label{FreeFall}   
\end{eqnarray}
Equation (\ref{SchAmp}) is well-known Regge-Wheeler equation.
Boundary conditions for this equation in our problem must represent outgoing wave at spatial infinity and ingoing wave at the horizon \cite{Tiomno1972,Ruffini1972,ThorneMacdonald1}
\begin{subequations}\label{boundaryconditions}
	\begin{equation}
	\tilde{b}_{l}(r^*,\,\omega)\rightarrow \alpha(\omega)e^{i\omega r^*}\,,\text{ for }r^*\rightarrow +\infty\,;\label{boundaryconditions1}
	\end{equation}
	\begin{equation}
	\tilde{b}_{l}(r^*,\,\omega)\rightarrow \beta(\omega)e^{-i\omega r^*}\,,\text{ for }r^*\rightarrow -\infty\,.\label{boundaryconditions2}
	\end{equation}
\end{subequations}
Here $\alpha(\omega)$ and $\beta(\omega)$ are certain unknown functions.

Consider the homogeneous equation, corresponding to (\ref{SchAmp})
\begin{equation}\label{Hom}
y_A''+\left[\omega^2-U(r^*)\right]y_A=0\,.
\end{equation}
Here capital Latin indices $A,\,B,\,...$ run from 1 to 2, so that $y_1(r^*)$ and $y_2(r^*)$ are two linear independent solutions of (\ref{Hom}). They can be expressed through Heun functions, see, e.g. \cite{Fiziev} and \cite{2022arXiv221104544K}.  We choose these solutions in such a way that solution $y_1(r^*)$ satisfies the boundary condition at the event horizon of the black hole (\ref{boundaryconditions2}), while solution $y_2(r^*)$ satisfies the boundary condition at spatial infinity (\ref{boundaryconditions1}). Then, using Green's function approach, the solution of (\ref{SchAmp}) satisfying the boundary conditions (\ref{boundaryconditions}) can be expressed  as \cite{1972NCimL...3..211R} 
\begin{eqnarray}
&&\tilde{b}_{l}(r^*,\,\omega)=\frac{1}{W(r^*)}\nonumber\\
&&\times\left[y_1(r^*)\int\limits^{+\infty}_{r^*}y_2(x)a(x)e^{i\omega T(x)}\mathrm{d}x\right.\nonumber\\
&&+\left. y_2(r^*)\int\limits^{r^*}_{-\infty}y_1(x)a(x)e^{i\omega T(x)}\mathrm{d}x\right]\,,\label{generalS}
\end{eqnarray}
where $W(r^*)$ is Wronskian
\begin{equation}\label{Wronskian}
W(r^*)=y_1(r^*)y_2'(r^*)-y_2(r^*)y_1'(r^*)\,.   
\end{equation}
Further simplification of  (\ref{generalS}) is possible by expressing the functions $y_A(r^*)$ through their derivatives using (\ref{Hom}) and performing integration by parts. The result reads
\begin{subequations}
	\begin{eqnarray}
	&\tilde{b}_l(r^*,\omega)=\dfrac{q}{2\pi}\dfrac{2l+1}{l(l+1)}\dfrac{e^{i\omega T(r^{*})}}{i\omega}\nonumber\\
	&+\underbrace{\dfrac{y_2(r^{*})}{W}\dfrac{q}{2\pi}\dfrac{2l+1}{i\omega}\int\limits_{-\infty}^{r^{*}}U(x)y_1(x)e^{i\omega T}\mathrm{d}x}_{I}\,,\nonumber\\
	&\text{ for } l\geqslant 0\,;\label{lastintegral}\\
	&\tilde{b}_0(r^*,\omega)=\dfrac{q}{2\pi}\dfrac{e^{i\omega T(r^{*})}}{i\omega}\,.\label{b0result}
	\end{eqnarray}
\end{subequations}

\section{Monopole field}
\label{sec3}
Clearly, monopole component $l=0$ of the electromagnetic field is static and it does not contain wave part. For this component the boundary conditions of the form  (\ref{boundaryconditions}) cannot be used. Nevertheless, solution for $l=0$ can be obtained using the Stokes theorem, see, e.g. \cite{Stephani,Schouten}
\begin{eqnarray}
&&q=\int\limits_{\Sigma}j^i\mathrm{d}S_i=\frac{1}{4\pi}\int\limits_{\Sigma}F^{ij}{}_{;j}\mathrm{d}S_i=\frac{1}{4\pi}\oint\limits_{\sigma}F^{ij}\mathrm{d}\sigma_{ij}\,,\label{Gauss}
\end{eqnarray}
where $\Sigma$ is a space-like hypersurface, that can be chosen as $t=\mathrm{const}$, $\sigma$ is a closed surface in $S$, that can be choosen as sphere $t=\mathrm{const}$, $r=\mathrm{const}$. Let $(\underset{1}{\mathrm{d}}x^k,\,\underset{2}{\mathrm{d}}x^k)$ be infinitesimal coordinate basis in $\sigma$, $(\underset{1}{\mathrm{d}}x^k,\,\underset{2}{\mathrm{d}}x^k,\,\underset{3}{\mathrm{d}}x^k)$ be the corresponding basis in $\Sigma$ and $(\underset{1}{\mathrm{d}}x^k,\,\underset{2}{\mathrm{d}}x^k,\,\underset{3}{\mathrm{d}}x^k,\,\underset{4}{\mathrm{d}}x^k)$ be the basis in the whole space-time.  Then (see, e.g. \cite{Schouten})
\begin{eqnarray}
&&\mathrm{d}S_i=\varepsilon_{ijkl}\underset{1}{\mathrm{d}}x^j\underset{2}{\mathrm{d}}x^k\underset{3}{\mathrm{d}}x^l\,;\\
&&\mathrm{d}\sigma_{ij}=\varepsilon_{ijkl}\underset{1}{\mathrm{d}}x^k\underset{2}{\mathrm{d}}x^l\,,  \nonumber
\end{eqnarray}
where $\varepsilon_{ijkl}$ is Levi-Chivita pseudotensor. Due to the orthogonality of Legendre polynomials on a sphere, the last integral in (\ref{Gauss}) is zero for all multipole components of $F^{ij}$ apart from the monopole component. So we obtain
\begin{eqnarray}\label{b0}
&&b_0(r,t)=r^2 (F^{10})_0=\nonumber\\
&&\left\{
\begin{split}
&-q\,, \text{ when the charge is inside  the surface } \sigma\,,\\
&0\,, \text{ when the charge is outside the surface } \sigma\,.\label{F}
\end{split}
\right.
\end{eqnarray}
Here $(F^{10})_0$ denotes the monopole term in multipole expansion of $F^{10}$. 

It can be shown that this result is in agreement with previous formulas. Indeed, one can start from the general solution for $b_0$ that can be obtained by adding the general solution of homogeneous equation (\ref{Hom}) (in our case $U(r^*)=0$ because of  $l=0$) to a particular solution of equation (\ref{SchAmp})  in the form (\ref{b0result}) as follows
\begin{equation}\label{0lastintegral}
\tilde{b}_0(\omega,r)=\frac{q}{2\pi}\frac{e^{i\omega T(r^{*})}}{i\omega}+Ae^{i\omega r^*}+Be^{-i\omega r^*}\,.  
\end{equation}
Here $A$ and $B$ are certain constants. Assuming that external electromagnetic field is absent and the particle starts at spatial infinity $r\rightarrow\infty$ without initial velocity ($v_{\infty}=0$), the choice should be $A=0$, $B=q/(2\pi i\omega)$. Then inverse Fourier transform gives
\begin{eqnarray}
&&\int\limits_{-\infty}^{+\infty}\tilde{b}_0(\omega,r)e^{-i\omega t}\mathrm{d}\omega\nonumber\\
&&=\frac{q}{2}\sgn\left(T(r^*)-t\right)+\frac{q}{2}\sgn\left(-r^*-t\right)\,,\label{b0fun}
\end{eqnarray}
where the function $\sgn(x)$ is equal to $+1$ for $x>0$ and it is equal to $-1$ if $x<0$. We are interesting only in the cases when particle moves in the vicinity of black hole and influence of black hole is not negligible. Therefore Without loss of generality one can assume $t>0$. For realistic physical situation the observer is outside of the photon sphere ($r>3M$). For this case it is obvious that the solution (\ref{b0fun}) coincides with (\ref{F}) (we take into account (\ref{Fl})). Note, that the result of this section, in particular, formula (\ref{0lastintegral}) are true for any function $T(r^*)$. 

\section{The case $l\geqslant 1$}
\label{sec4}

The goal of the present section is to obtain asymptotic solution for Fourier coefficients $b(r,t)$ when radial coordinate of the particle $r\rightarrow 2M$ and $l\geqslant 1$. This solution can be obtained by applying inverse Fourier transform to (\ref{lastintegral}). For this purpose it is necessary to know analytic expression for the integral in term $I$ of (\ref{lastintegral}) that can be obtained by using approximate solution $y_1$ of homogeneous equation (\ref{Hom}). 

\begin{figure}
	\centering
	\includegraphics[angle=0, width=\columnwidth]{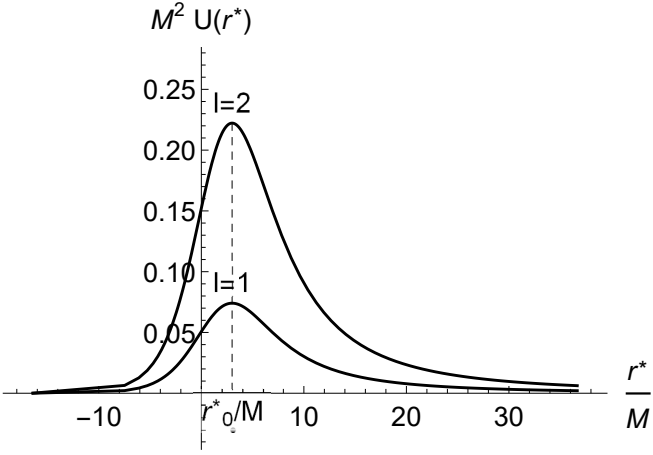}
	\caption{Potential barrier of the Regge-Wheeler equation for multipole moments $l=1$ and $l=2$.} \label{barrier}
\end{figure}

It is well known, see, e.g. \cite{1973blho.conf..451R}, that equation (\ref{Hom}) for the variable $r^*$ mathematically coincides with stationary Schrodinger equation for a particle in a potential barrier $U(r^*)$ relative to Cartesian variable $x=r^*$. This barrier is represented in Fig. \ref{barrier}. 
It has a maximum for $r^*=r^*_0\equiv 3M+2M\ln{(1/2)}$. The simplest way to obtain approximate analytic solution of this equation is to represent such a barrier with the Dirac delta function located in the maximum of the potential
\begin{equation}
U(r^*)=U_0\delta(r^*-r^*_0)\,.
\end{equation}
Here $U_0$ is a constant. Choose two solutions of (\ref{Hom}) in the following form
\begin{eqnarray}\label{y1}
y_1(r^{*})=\left\{
\begin{split}
&e^{-i\omega r^{*}}\,,\quad r<3M\,,\\
&Ce^{-i\omega r^{*}}+De^{i\omega r^{*}}\,;\quad r>3M\,;
\end{split}
\right.
\end{eqnarray}	
and
\begin{eqnarray}\label{y2}
y_2(r^{*})=\left\{
\begin{split}
&Ce^{i\omega r^{*}}+De^{-i\omega r^{*}}\,,\quad r<3M\,;\\
&e^{i\omega r^{*}}\,,\quad r>3M\,,
\end{split}
\right.
\end{eqnarray}
where coefficients have the following form (see, e. g. \cite{Flugge})
\begin{eqnarray}
&C=1-\frac{U_0}{2i\omega}\,;\\
&D=\frac{D}{2i\omega}e^{-2i\omega\cdot 3M^{*}}\,;\\
&W=2i\omega C=2i\omega-U_0\,.
\end{eqnarray}

Due to the asymptotic form of (\ref{y1}) and (\ref{y2}) for $r^*\rightarrow\pm\infty$ it follows from (\ref{SchAmp}) that boundary conditions (\ref{boundaryconditions2}), (\ref{boundaryconditions1}) are satisfied for the functions $y_1(r^*)$ and $y_2(r^*)$ respectively. Term $I$ in (\ref{lastintegral}) has the following form
\begin{eqnarray}
&&\dfrac{q}{2\pi}\dfrac{2l+1}{l(l+1)}\dfrac{U_0}{i\omega}\dfrac{e^{i\omega r^{*}}}{2i\omega-U_0}\int\limits_{-\infty}^{r^{*}}\delta(x-r^*_0)
e^{-ix\omega}e^{i\omega T(x)}\mathrm{d}x\nonumber\\
&&=\dfrac{q}{2\pi}\dfrac{2l+1}{l(l+1)}\dfrac{U_0}{i\omega}\frac{e^{i\omega (r^{*}+T(r^*_0)-r^*_0)}}{2i\omega-U_0}\,.\label{bFurie}
\end{eqnarray}
Inverse Fourier transform gives
\begin{equation}\label{bInt}
U_0\frac{q}{2\pi}\frac{2l+1}{l(l+1)}\int\limits_{-\infty}^{+\infty}\frac{e^{i\omega (r^{*}+T(r^*_0)-r^*_0-t)}}{i\omega (2i\omega-U_0)}\mathrm{d}\omega\,.
\end{equation}
This integral (\ref{bInt}) can be performed using Cauchy's residue theorem. The contour of integration depends on the sign of the factor under the exponent. Due to this consider two cases. For $r^{*}+T(r^*_0)-r^*_0-t>0$ one has
\begin{eqnarray}
&&b_l=\pi i \underset{\omega=0}{\res}\left[U_0\frac{q}{2\pi}\frac{2l+1}{l(l+1)}\frac{e^{i\omega (r^{*}+T(r^*_0)-r^*_0-t)}}{i\omega (2i\omega-U_0)}\right]\nonumber\\
&&=-\frac{\rho}{2}\frac{2l+1}{l(l+1)}\,.\label{bInt1}
\end{eqnarray}
If $r^{*}+T(r^*_0)-r^*_0-t<0$ one has
\begin{eqnarray}
&&b_l=-\pi i \underset{\omega=0}{\res}\left[U_0\frac{q}{2\pi}\frac{2l+1}{l(l+1)}\frac{e^{i\omega (r^{*}+T(r^*_0)-r^*_0-t)}}{i\omega (2i\omega-U_0)}\right]\nonumber\\
&&-2\pi i \underset{\omega=-iU_0/2}{\res}\left[U_0\frac{\rho}{2\pi}\frac{2l+1}{l(l+1)}\frac{e^{i\omega (r^{*}+T(r^*_0)-r^*_0-t)}}{i\omega (2i\omega-U_0)}\right]\nonumber\\
&&=\frac{q}{2}\frac{2l+1}{l(l+1)}-\rho\frac{2l+1}{l(l+1)}e^{\frac{U_0}{2}\omega (r^{*}+T(r^*_0)-r^*_0-t)}\,.\label{bInt2}
\end{eqnarray}
Substituting (\ref{bInt1}) and (\ref{bInt2}) into (\ref{lastintegral}), finally we get
\begin{eqnarray}\label{bdelta}
b_l(r(r^*),t)=\left\{
\begin{split}
& -\rho\frac{2l+1}{l(l+1)}\,, \\
& \text{ if } T(r^{*})<t<r^{*}+T(r^*)-r^*_0\,;\\
& -\rho\frac{2l+1}{l(l+1)}e^{\frac{U_0}{2}\omega (r^{*}+T(r^*_0)-r^*_0-t)}\,,\\ 
& \text{ if } t>r^{*}+T(r^*_0)-r^*_0\,.
\end{split}
\right.
\end{eqnarray}
It follows from (\ref{bdelta}) that the value $b_l$ for the particle that is approaching asymptotically the black hole event horizon ($t\rightarrow +\infty$) tends to zero exponentially. In order to determine all components of electromagnetic field we use the Coulomb gauge $A^i{}_{;i}=0$, $A_4=0$. Substituting this in equation (\ref{anzats}) we find
\begin{eqnarray}
&&0=\sum\limits_{l=0}^{\infty}\left[\pd{}{r}\left(h_l(r,t)\left(1-\frac{2M}{r}\right)\right)+\frac{2}{r}\left(1-\frac{2M}{r}\right)\right.\nonumber\\
&&\left.-\frac{l(l+1)}{r^2}k_l(r,t)
\right]P_l(\cos\theta)\,.\nonumber\\\label{Fto0}
\end{eqnarray}

Taking into account that $h_l(r,t)$ tends to zero exponentially, it follows from (\ref{Fto0}) that on the black hole horizon also $k_l(r,t)$ and consequently all components of electromagnetic field tend to zero for $t\rightarrow +\infty$ exponentially as $\exp\left[\frac{1}{2}U_0\omega (r^{*}+T(r^*_0)-r^*_0-t)\right]$.

Note that this result is independent on the function $T(r^*)$. It is valid also when external electromagnetic field as well as radiation reaction are present.

\section{Rectangular potential barrier}
\label{sec5}
\begin{figure}
	\centering
	\includegraphics[angle=0, width=\columnwidth]{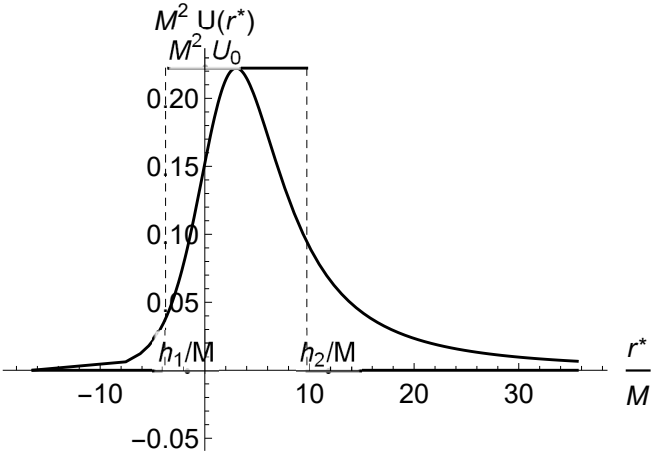}
	\caption{The approximation of the Regge-Wheeler potential by a rectangular potential.} \label{Rectangular}
\end{figure}

In this section we consider another approximation for the potential barrier $U(r^*)$ in the form of rectangular barrier. It has three parameters: height $U_0$ and positions of the walls $h_1$ and $h_2$, see Fig. \ref{Rectangular}.
Ingoing and outgoing solutions ($y_1$ and $y_2$ respectively) have the following form
\begin{eqnarray}\label{y1R}
y_1(r^*)=\left\{
\begin{split}
& e^{-i\omega r^*}, \text{ for }r^*<h_1\,; \\
& Me^{i\sqrt{\omega^2-U_0}(r^*-h_1)}+Ne^{-i\sqrt{\omega^2-U_0}(r^*-h_1)}, \\
&\text{ for }h_1<r^*<h_2\,;\\
&Ce^{i\omega(r^*-h_2)}+De^{-i\omega(r^*-h_2)}, \text{ for }r^*>h_2\,;\\ 
\end{split}
\right.
\end{eqnarray}
\begin{eqnarray}\label{y2R}
y_2(r^*)=\left\{
\begin{split}
& De^{i\omega(r^*-h_1)}+Ce^{-i\omega(r^*-h_1)}, \text{ for }r^*<h_1\,; \\
& Ne^{i\sqrt{\omega^2-U_0}(r^*-h_1)}+Me^{-i\sqrt{\omega^2-U_0}(r^*-h_1)}, \\
&\text{ for }h_1<r^*<h_2\,;\\
&e^{i\omega r^*}, \text{ for }r^*>h_2\,.\\ 
\end{split}
\right.
\end{eqnarray}
Here coefficients $M,\,N,\,C,\,D$ depend only on the frequency parameter $\omega$ and on the parameters of the barrier $h_1,\,h_2,\,U_0$. Exact expressions for these functions can be found in textbooks on quantum mechanics (see, e. g. \cite{Flugge}). The Wronskian (\ref{Wronskian}) can be computed using (\ref{y1R}) and (\ref{y2R}) with the result
\begin{equation}\label{W}
W(\omega)=2i\omega D\,.
\end{equation}
After inverse Fourier transform the term $I$ in (\ref{lastintegral}) is
\begin{equation}\label{inversetransform}
\frac{y_2(r^{*})}{W}\frac{q}{2\pi}\frac{2l+1}{i\omega}\int\limits_{-\infty}^{r^{*}}U(x)y_1(x)e^{i\omega T(x)-t}\mathrm{d}x\,.
\end{equation}
For rectangular potential the solutions (\ref{y1R}) and (\ref{y2R}) of homogeneous equation and Wronskian (\ref{W}) this integral becomes
\begin{widetext}
	\begin{eqnarray}
	&U_0\dfrac{q}{2\pi}\dfrac{2l+1}{l(l+1)}\int\limits_{h1}^{h2}\mathrm{d}x
	\int\limits_{-\infty}^{\infty}\mathrm{d}\omega \dfrac{e^{i\omega(r^*-h_2+T(x)-t)}}{i\omega}\times\nonumber\\
	&\dfrac{\sqrt{\omega^2-U_0}\cos{(\sqrt{\omega^2-U_0}(x-h_1))}-i\omega\sin{(\sqrt{\omega^2-U_0}(x-h_1))}}{i\omega\sqrt{\omega^2-U_0}\cos{(\sqrt{\omega^2-U_0}(h_2-h_1))}
		+(2\omega^2-U_0)\sin{(\sqrt{\omega^2-U_0}(h_2-h_1))}}\,.\nonumber\\
	\label{integral}
	\end{eqnarray}
\end{widetext}
We calculate the integral (\ref{integral}) with respect to $\omega$, by using well-known asymptotic formula (see, e. g. \cite{Vladimirov})
\begin{eqnarray}
&\lim\limits_{\lambda\rightarrow +\infty}\left(\mathrm{v.p.} \int\limits_{-\infty}^{\infty}f(\omega)\dfrac{e^{-i\lambda\omega}}{i\omega}\mathrm{d}\omega\right)= -\pi f(0)\,,  
\end{eqnarray}
Here $f(\omega)$ is an arbitrary analytical function for $\omega\in(-\infty;\,+\infty)$. Note that this formula also valid in the case of non-analytical $f(\omega)$ that has only branch points where $f(\omega)$ is finite.  Symbol v.p. denotes main value of the integral relative to the singular point $\omega=0$. In the case of (\ref{integral}) $\lambda=-(r^*-h_2+T(x)-t)$ and it diverges for $t\rightarrow\infty$. Then, the limiting value for (\ref{integral}) has the form
\begin{eqnarray}
&\dfrac{q}{2}\dfrac{2l+1}{l(l+1)}\int\limits_{h_1}^{h_2}\sqrt{-U_0}\dfrac{\cos{(\sqrt{-U_0}(x-h_1))}}{\sin{(\sqrt{-U_0}(h_2-h_1))}}\mathrm{d}x\nonumber\\
&=\dfrac{q}{2}\dfrac{2l+1}{l(l+1)}\,.
\end{eqnarray}
Performing inverse Fourier transform of all expression (\ref{lastintegral}) we find
\begin{eqnarray}
&\lim\limits_{t\rightarrow\infty}b_l(r^*,t)=\lim\limits_{t\rightarrow\infty}\dfrac{q}{2}\dfrac{2l+1}{l(l+1)}\sgn{(T(r^*)-t)}\nonumber\\
&+\dfrac{q}{2}\dfrac{2l+1}{l(l+1)}=0\,.\label{result}
\end{eqnarray}
This result confirms that electromagnetic field is asymptotically ($t\rightarrow\infty$) spherically symmetric in this approximation. Note, that sufficient condition for this: $\lambda\rightarrow\infty$ physically means, that particle should cross the photon sphere ($r=3M$ is Schwarzschild coordinates). 

\section{Discussion}
The main result of this work is the demonstration that the infall of a charge on a Schwarzschild black hole results for asymptotically large time in a spherically symmetric electrostatic field in the framework of the  \emph{dynamical} problem in classical general relativity. In the literature (see, e.\,g. \cite{ThorneMacdonald,1973blho.conf..451R}) the electromagnetic field of the charged particle was considered when the particle is kept at rest with respect to the black hole, which is unphysical. In the present work we solve the dynamical problem when the source is radially moving near the black hole. Our result is obtained analytically by approximating the tunnelling barrier of the Regge-Wheeler radial equation by Dirac delta-function as well as by a rectangular potential. This implies that in the region of spacetime accessible to measurements by the distant observer, the end result of the infall of a charged particle into a Schwarzschild black hole is the spherically symmetric solution, \emph{indistinguishable from the Reissner–Nordstr\"om solution}. This is true, even if for external observer the charge never reaches the horizon of the black hole. It is interesting that these conclusions do not depend on parameters of the rectangular potential barrier $U_0,\,h_1,\,h_2$ or delta-function $U_0$. 

We note that these results were obtained for the case of charge moving along radial geodesic in Schwarzschild coordinates. However all expressions are still valid for any monotonic function $T(r^*)$ ($T'(r^*)<0$). In particular, our results are valid in the case of radial motion in the presence of external electromagnetic field and when electromagnetic reaction of the particle is taken into account. Then we can say that in this case part of electromagnetic field generated by the particle is spherically symmetric when the particle approaches event horizon and other multipole moments are coming only from external electromagnetic field. For example, our results apply as well to the case when charge accretion occurs on the symmetry axis of a magnetized Schwarzschild black hole.

It is necessary to add that our results cannot be considered as rigorous mathematical proof that the electromagnetic field of a charged particle, falling into black hole is spherically symmetric for $t\rightarrow\infty$. This is due to approximating character of obtained analytical formulas. 

Given the linearity of Maxwell equations the same results also
holds for charged black holes as well as for accretion of charges with different signs. In particular, it follows that when black hole is accretring neutral plasma no charge separation could occur, hence no charge is induced on the black hole. This is particularly relevant in models of cosmic transients where plasma around black holes is considered as the source of radiation 
\cite{2018PhRvD..98l3002L,2019PhRvL.122c5101P,2022ApJ...929...56R,2023PhRvD.107d4055A,King,Komissarov}

\begin{acknowledgments}
	The work was supported by BRFFR Foundation in the framework of the F21ICR BRFFR-ICRANet project. GV would like to express his gratitude to prof. Vladimir Belinski for numerous discussions on this topic.
\end{acknowledgments}

%

\end{document}